\documentclass[twocolumn,aps]{revtex4}
\usepackage{graphicx}

\begin{document}
\title{Picture Invariance in Quantum Optics}

\author{Won-Young Hwang$^{*}$}

\affiliation{Department of Physics Education, Chonnam National
University, Gwangju 500-757, Republic of Korea}

\begin{abstract}
We clarify the controversy over the coherent-state (CS) versus the
number-state (NS) pictures in quantum optics. The NS picture is
equivalent to the CS picture, as long as the phases $\phi$ in the
laser fields are randomly distributed, as M\o lmer argues [\pra {\bf
55}, 3195 (1997)]. However, the claim by Rudolph and Sanders [Phys.
Rev. Lett. {\bf 87}, 077903 (2001)] has a few gaps. First, they make
an assumption that is not necessarily true in the calculation of a
density operator involved with a two-mode squeezed state. We show
that there exists entanglement in the density operator without
defying the assumption that phases are randomly distributed.
Moreover, using a concept of picture-invariance, we argue that it is
not that criteria for quantum teleportation are not satisfied. We
discuss an analogy between the controversy on the CS versus NS
pictures to that on the heliocentric versus geocentric pictures.

\end{abstract}

\maketitle
%%%%%%%%%%%%%%%%%%%%%%%%%%%%%%%%%%%%
\section{Introduction}
%%%%%%%%%%%%%%%%%%%%%%%%%%%%%%%%%%%%
Besides continuing controversies over its implication on our
understanding of the physical world
\cite{Ein35,Sel88,Bel64,Boh93,Hol93}, quantum mechanics is still
revealing its hidden aspects, now in a form of quantum information
processing \cite{Nie00}. The coherent-state (CS) picture has been
successful in describing quantum optical phenomena
\cite{Gla63,Yue76,Scu97}. However, the fact that a picture is
successful does not exclude the possibility of other pictures.
Indeed, the CS picture is not the only choice \cite{Mol97}.

Quantum teleportation (QT) \cite{Ben93} is an interesting ingredient
of quantum information processing: Utilizing the nonlocality of
Einstein-Podolsky-Rosen pairs of \cite{Ein35,Sel88,Bel64} quantum
bits (qubits), QT enables us to do a task whose result is equivalent
to actual transport of qubits. Recently, a quantum optical
experiment was performed by Furusawa {\em et al.} \cite{Fur98} to
demonstrate a continuous-variable version \cite{Vai94,Bra98} of QT.
However, Rudolph and Sanders (RS) \cite{Rud01} criticized the
experiment in Ref. 15 by extending the argument of M\o lmer
\cite{Mol97}. RS were then criticized by van Enk and Fuchs
\cite{Enk02}. Sanders {\em et al.} again supported Ref. 18 in a
subsequent work \cite{San03}. Besides those, a few authors have
joined the controversy \cite{Nem04,Wis03,Smo04}. In fact, Refs. 21
and 22 give criticisms that are quite similar to ours. However, our
viewpoint is not the same as theirs \cite{Nem04,Wis03}, especially
in that we discuss a concept of `picture' in connection with the
controversy.

The purpose of this paper is to clarify the controversy. The
number-state (NS) picture is equivalent to the CS picture in the
sense that the NS picture can also explain quantum optical
phenomena, as M\o lmer argues \cite{Mol97}. However, the arguments
of RS have a few gaps. First there is a technical gap in their
interpretation of the experiment, as pointed out by Yuen
\cite{Yue02}. Next, although we cannot fix our picture, we can
define certain properties that are invariant for transformation
between various pictures, including the CS and the NS pictures.
(This is in analogy with the fact that although we cannot fix our
reference frame, we have some facts, e.g., 'two objects are in
contact', that are invariant to transformation of the reference
frame in classical mechanics.) Using the concept of
picture-invariance, we can see that the usual interpretation using
the CS picture is meaningful enough. This paper is organized as
follows: First, we describe the NS picture. Then, we point out the
technical gap in the argument of RS. Next, we introduce
picture-invariance. Then, we argue, with the picture-invariance, how
RS's claim is not valid. Then, we discuss an analogy between the
controversy on CS versus NS pictures to that on the heliocentric
versus geocentric pictures, and we conclude.
%%%%%%%%%%%%%%%%%%%%%%%%%%%%%%%%%%%%
\section{Number state picture}
%%%%%%%%%%%%%%%%%%%%%%%%%%%%%%%%%%%%
Let us discuss a fact about an absolute phase $\phi$ of the laser
field: Even if we can perform a real phase measurement, the absolute
value of the phase has no meaning \cite{Enk05}. The meaning becomes
clear when we consider an analogy between the phase value and
spatial coordinate value. Even if we can measure the distance
$\Delta x= x_1 -x_2$ between two points $x_1$ and $x_2$ in space, it
is meaningless to argue what absolute value should be assigned to a
point, e.g., $x_1$. In other words, we can assign any value to a
point. In this sense only, we cannot measure or determine the
absolute phase of a laser field (Proposition-0).

However, an actual laser field is not in a pure coherent state
$|\alpha e^{i\phi} \rangle = \exp(-\alpha^2/2) \sum_{n=0}^\infty
(\alpha^n e^{in\phi}/\sqrt{n!})| n \rangle$, but in a mixed state
\begin{eqnarray}
\label{A} \rho_L &=& \int_0^{2\pi} \frac{d\phi}{2\pi} |\alpha
e^{i\phi} \rangle
                \langle \alpha e^{i\phi}|
                      \\
        &=& e^{-\alpha^2} \sum_{n=0}^\infty
 \frac{\alpha^{2n}}
        {n!} | n \rangle \langle n |.
\end{eqnarray}
The reason is the following: Values of the phase $\phi$ in laser
fields are randomly distributed (Proposition-1) \cite{Mol97}. As a
principle of quantum mechanics, however, we cannot distinguish two
different decompositions of the same density operator \cite{Nie00}.
Therefore, we may well choose any decomposition for the laser field
$\rho_L $ in describing quantum optical phenomena involved with
$\rho_L $. For example, it can be either a decomposition in Eq. (1)
(CS picture) or that in Eq. (2) (NS picture). As long as
Proposition-1 and quantum mechanics are correct, the above argument
by M\o lmer \cite{Mol97} cannot be incorrect.
%%%%%%%%%%%%%%%%%%%%%%%%%%%%%%%%%%%%
\section{A gap in the arguments of Rudolph and Sanders}
%%%%%%%%%%%%%%%%%%%%%%%%%%%%%%%%%%%%
Let us be reminded of the argument by RS \cite{Rud01}. If a pure
coherent state $|\alpha e^{i\phi} \rangle$ is used to pump nonlinear
crystals, the two-mode squeezed state
\begin{equation}
\label{C} |\eta e^{i \phi}\rangle = \sqrt{1-\eta^2}
\sum_{n=0}^{\infty} \eta^n e^{i n \phi} |nn\rangle
\end{equation}
is generated. Here, the phase $\phi$ of the pumping state is
transcribed into that of a generated state. Since the
phase-randomized state $\rho_L$ is used to pump, the generated state
$\rho_S$ is also phase randomized,
\begin{equation}
\label{D} \rho_S = \int_0^{2\pi} \frac{d\phi}{2\pi} \, \vert \eta
e^{i\phi} \rangle
                \langle \eta e^{i\phi} \vert
                \nonumber       \\
    = \left( 1 - \eta^2 \right) \sum_{n=0}^\infty
        \eta^{2n} \vert n \, n \rangle \langle n \, n \vert.
\end{equation}
A reasonable definition of separability is that as long as a density
operator $\rho$ is separable in any decomposition, it is separable
\cite{Nie00}. Thus, the phase-randomized two-mode squeezed state
$\rho_S$ is separable. RS claim that there is no entanglement in
places where entanglement is supposed to be in normal QT. In their
criticism for RS, van Enk and Fuchs \cite{Enk02} make a claim that
there is entanglement in the state. However, they assume an
experiment that measures the phase of a laser field. The problem is
that the experiment has not yet been realized although the
possibility is not excluded in principle. van Enk and Fuchs claim
that whether we perform the experiment or not, the fact that there
exist entanglement in the state of Eq. (12) of Ref. 19 does not
change. However, this is not the case. Let us consider a similar
illustrating example where Alice and Bob share a state
\begin{eqnarray}
\label{G}  |\Psi\rangle= \frac{1}{2}(|0\rangle |\phi^+\rangle +
|1\rangle |\phi^-\rangle + |2\rangle |\psi^+\rangle + |3\rangle
|\psi^-\rangle)
\end{eqnarray}
Here, $|0\rangle, |1\rangle, |2\rangle$, and $|3\rangle$ are
normalized and orthogonal states, and, for example, $|0\rangle
|\phi^+\rangle$ denotes $|0\rangle_A |\phi^+\rangle_{AB}$ where $A$
and $B$ denote Alice and Bob, respectively. Alice and Bob are
assumed to be remotely separated as usual. Bell states are given by
$ |\phi^{\pm}\rangle_{AB}= (1/\sqrt{2})(|0\rangle_A |0\rangle_B \pm
|1\rangle_A |1\rangle_B)$ and $ |\psi^{\pm}\rangle_{AB}=
(1/\sqrt{2})(|0\rangle_A |1\rangle_B \pm |1\rangle_A |0\rangle_B)$.
A measurement that distinguishes the four states $\{|0\rangle,
|1\rangle, |2\rangle, |3\rangle \}$ would reduce the state
$|\Psi\rangle$ to one of the Bell states, and the outcome of the
measurement would identify which Bell state it is. Therefore, if
Alice can perform the measurement, it amounts to their sharing a
Bell state. However, if Alice has no way of performing the
measurement by a reason, e.g., because she lost the first qubit, the
second and the third qubits they share no longer constitute a Bell
state, but constitute a purely separable state described by
\begin{eqnarray}
\label{G-2} \frac{1}{4}(|\phi^+\rangle \langle \phi^+|+
|\phi^-\rangle \langle \phi^-|+|\psi^+\rangle \langle
\psi^+|+|\psi^-\rangle
\langle \psi^-|)&&\\
\label{H}=\frac{1}{4}\vspace{1mm} I = \frac{1}{4}(|00\rangle \langle
00|+ |01\rangle \langle 01| + |10\rangle \langle 10|+ |11\rangle
\langle 11|).&&
\end{eqnarray}
Here, $I$ is an identity operator. One might argue that an objective
property like entanglement must not depend on subjective knowledge.
However, it is one of the most interesting facts in quantum
information that entanglement depends on subjective knowledge. In
the same way, there is no entanglement in the state of Eq. (12) of
Ref. 19 before the phase measurement is realized.

What we show here is that even without the phase measurement there
remains entanglement in the usual two-mode squeezed state. The
problem in RS's argument is that in their interpretation of the
experiment, they make an assumption that is not necessarily true: A
laser used to pump a nonlinear crystal is reset each time when a
pair of photon pulses is generated (Prescription-1). However, {\it
Prescription-1 is not followed in a real experiment.} Once the
pumping laser is turned on, it is used for a while. However, a
coherent state $|\alpha e^{i\phi} \rangle$ can be split into copies
of the same phase in $m$ different modes,
\begin{equation}
\label{E} |\alpha e^{i\phi} \rangle \rightarrow
|\frac{\alpha}{\sqrt{m}} \hspace{1mm} e^{i\phi} \rangle^{\otimes m},
\end{equation}
where $m$ is a positive integer and $\alpha$, $\beta$ are real
numbers. For the case of spatial modes, the result in Eq. (\ref{E})
can be obtained by simple beam splitters and phase-shifters, as is
well known \cite{Scu99}. For the case of temporal modes, the result
in Eq. (\ref{E}) is obtained if we keep using the same laser beam
for a duration of time \cite{Enk02}. In either case, each state
$|\frac{\alpha}{\sqrt{m}} \hspace{1mm} e^{i\phi} \rangle$ in $m$
different modes is used to pump a nonlinear crystal. Then, the
phases of the two-mode squeezed state are the same. The
corresponding density operator is, therefore, not $\rho_S$, but
\begin{equation}
\label{F} \rho_T = \int_0^{2\pi} \frac{d\phi}{2\pi} \,
|\frac{\eta}{\sqrt{m}} e^{i\phi} \rangle ^{\otimes m}
                \langle \frac{\eta}{\sqrt{m}} e^{i\phi}| ^{\otimes
                m}.
\end{equation}
Note that the states $\rho_S$ and $\rho_T$ are different. Although
the state $\rho_S$ has no entanglement, the state $\rho_T$ has
entanglement.
%%%%%%%%%%%%%%%%%%%%%%%%%%%%%%%%%%%%
\section{Picture invariance and existence of entanglement}
%%%%%%%%%%%%%%%%%%%%%%%%%%%%%%%%%%%%
Before showing the existence of entanglement in the state $\rho_T$,
let us discuss picture-invariance. Picture-invariant quantities or
properties are those that do not depend on any particular
decomposition of a certain density operator. One example is the
density operator itself. Another example is existence of
entanglement or non-separability. Recall the definition of
non-separability: If a density operator is not separable for any
decomposition of the density operator, it is non-separable
\cite{Nie00}. Measures of entanglement, e.g., entanglement of
formation \cite{Nie00}, are also picture-invariant. Note that we may
well use any particular decomposition in estimating
picture-invariant quantities or properties. In other words, whatever
picture we use, we get the same result for picture-invariant
quantities or properties.

Now let us show that the state $\rho_T$ has entanglement. That may
be done by proving that a measure of entanglement for the state
$\rho_T$ has a non-vanishing value. However, our method is to show
that the state $\rho_T$ can violate Bell's inequality
\cite{Bel64,Sel88}. It is known  \cite{Ban98,Jeo00} that the
two-mode squeezed state $|\eta e^{i\phi} \rangle ^{\otimes m} $ can
violate the Bell's inequality by measuring even and odd parities if
all instruments share a single reference beam and if Prescription-1
is not followed. Here, the problem is that a beam (the reference
beam), besides the beams corresponding to the state $\rho_T$, is
shared by Alice and Bob. Therefore, even if the Bell's inequality is
violated, one cannot say that it is solely due to the state
$\rho_T$. However, we consider an experiment where Alice and Bob do
not share any beam except for the ones corresponding to the state
$\rho_T$. Even in this case, however, we can see that it can violate
Bell's inequality: Let us assume that the Bell's inequality is
violated in a specific case where $\phi=p$, $\alpha= a$, and
$\beta=b$ in Eq. (6) of Ref. 28. We repeat many times the experiment
for the state $\rho_T$. In each experiment, $m$ pairs of photon
pulses are measured. $m$ is large enough to give statistical
confirmation. The phase is randomly given in each experiment because
Alice and Bob do not share a reference beam. When the phases are far
from the desired values, $p$, $a$, and $b$, Bell's inequality may
not be violated. However, when the phases happen to be close to the
desired ones, Bell's inequality is violated. The result is that
Bell's inequality is violated for $m$ pairs with a fixed success
probability $P$. One might say that in this case, we cannot say that
Bell's inequality is violated because we have selected certain
samples that violate Bell's inequality. However, that is not the
case because the number $m$ can be large enough with a fixed success
probability $P$. Assume that we simply post-select samples that
violate Bell's inequality in a row from a larger sample that does
not violate Bell's inequality. In this case, the success probability
$P$ should exponentially decrease with the size $m$ of the sample
that violate Bell's inequality. In the above case, however, the
success probability $P$ is fixed. Therefore, we can say that the
state $\rho_T$ is nonlocal and, thus, has entanglement.

We note that nonlocality of a state is a picture-invariant property.
Whatever picture we choose, we will get the same result for a given
state. Therefore, the above result obtained by the CS picture is
valid. Here, a picture is just a mathematical tool to get a
picture-invariant result.
%We note again that we are not saying that
%Proposition-A is incorrect, differently from van Enk and Fuchs
%\cite{Enk02}.

Let us now criticize RS's claim. Let us consider the three criteria
of QT \cite{Rud01}. (a) The state to be teleported is unknown to a
sender, Alice, and a receiver, Bob, and is supplied by an actual
third party, Victor. (b) Alice and Bob share only a nonlocal
entangled resource and a classical channel through which Alice
transmits her measurement results to Bob. (c) Entanglement should be
a verifiable resource. RS argue that none of the three criteria is
satisfied in the experiment by Furusawa {\em et al.} \cite{Fur98}.
As we have seen already, however, the criterion (c) is satisfied in
the experiment.

Next, let us consider the criterion (b). RS say that Alice and Bob
have extra entanglement via a pumping laser they share in the
experiment; thus, criterion (b) is not satisfied. In the NS picture,
apparently there is entanglement between the shared laser and light
beams split from it. However, in this case, the entanglement is not
a genuine one in the following sense: Let us consider again the
example of a mixed state of qubit-pairs whose density operator is
described by either Eq. (\ref{G-2}) or ({\ref{H}). In Eq. (\ref{H}),
however, the density operator is interpreted as a mixture of product
states while in Eq. (\ref{G-2}), the density operator is interpreted
as a mixture of highly entangled states. Separability is a
picture-invariant property. In other words, it is not reasonable
that a property like separability depends on the picture we choose.
Therefore, we can say that there is no entanglement in the density
operator $(1/4) \vspace{1mm} I$ because the density operator is
separable in a picture. In the same way, it is reasonable to say
that entanglement, between the shared laser and light beams split
from it, in the NS picture is not a real entanglement. One can
choose whatever picture he/she wants, but that is the case only when
a picture is used as a calculational tool. To summarize, because the
state of the system, except for the one corresponding to $\rho_T$,
is separable according to usual definition that is
picture-invariant, we can say that the criterion (b) is satisfied in
the experiment.

Let us interpret Smolin's discussion \cite{Smo04} on criterion (a)
with the concept of picture invariance. In the CS picture, the
states are separable. However, Victor rotates his beam with a phase
that is randomly chosen by him by using a phase shifter. The random
phase is what Alice and Bob do not know, which is a
picture-invariant quantity. What if we use the NS picture? Can the
information on Victor's action be transferred via (artificial)
entanglement in the NS picture? (RS argues that there are many
correlations in the two beams of Eq. (2) of Ref. 18. Also, they
implied that these correlations can be utilized in exchanging
information. However, what the correlations mean is just that the
states have {\it the same} phase in the CS picture before Victor's
rotating action.) However, a physical fact, like Alice and Bob not
being able to get information on Victor's rotating action via the
shared pumping laser, must be picture-invariant. This means that
even if there are entanglements in the NS picture, after careful
considerations, those entanglements will turn out not to be useful
for transferring information from Victor to Alice and Bob.
Therefore, we can say that the criterion (a) is also satisfied in
the experiment.
%%%%%%%%%%%%%%%%%%%%%%%%%%%%%%%%%%%%
\section{Discussion and conclusion}
%%%%%%%%%%%%%%%%%%%%%%%%%%%%%%%%%%%%
It is interesting that the controversy on the CS versus NS pictures
has an analogy to the old one on the heliocentric versus geocentric
pictures on motions of planets in the solar system. Both pictures
work as a tool for calculating observed phenomena. We cannot say
that an interpretation is correct while the other one is incorrect.
However, as Ocam proposed, what is simple or convenient should be
preferred. In this sense, the CS picture is better than the NS
picture: Picture-invariant properties like separability manifest
themselves in the CS picture while picture-invariant properties
often disguise themselves in the NS pictures. We are not committing
the partition-ensemble-fallacy \cite{Kok00}: We are not talking
about picture-specific ones, but talking about picture-invariant
ones.

%For works by \cite{Nem04,Wis03}, we have the same opinion as that of
%Smolin \cite{Smo04} that they have flaw in that the ensembles they
%identified can be plainly distinguished.

In conclusion, the NS picture is equivalent to the CS picture as M\o
lmer argues \cite{Mol97}, as long as Proposition-1 is correct.
However, claim by RS \cite{Rud01} has a few gaps. First, they make
an assumption that is not necessarily true in the calculation of a
density operator involved with two-mode squeezed state. We showed
that there exists entanglement in the density operator without
defying the Proposition-1. Moreover, using the concept of
picture-invariance that we introduced, we argued that two criteria
for QT are also satisfied. Then, we discussed an analogy between the
controversy on the CS versus NS pictures to that on the geocentric
versus heliocentric pictures. We argued why we were not committing
the partition-ensemble-fallacy \cite{Kok00}.

\acknowledgments This work was supported by the Korea Science and
Engineering Foundation (R01-2006-000-10354-0). I thank very much
Horace Yuen and Ranjith Nair for their enlightening discussions and
helpful comments. I thank very much S. J. van Enk for helpful
comments on the first version of the manuscript.

%%%%%%%%%%%%%%%%%%%%%%%%%%%%%%%%%%%%


\begin{references}
\bibitem[*]{email}
Email address: wyhwang@chonnam.ac.kr
\bibitem{Ein35} A. Einstein, B. Podolsky and N. Rosen,
                    Phys. Rev. {\bf47}, 777 (1935).
\bibitem{Sel88} F. Selleri, ed.,{\it Quantum Mechanics versus Local Realism.
                The Einstein-Podolsky-Rosen Paradox}
                (Plenum, New York, 1988).
\bibitem{Bel64} J. S. Bell, Physics {\bf1}, 195 (1964),
                 reprinted in
                  {\it Speakable and Unspeakable in Quantum
                  Mechanics} (Cambridge University Press, Cambridge, 1987).
\bibitem{Boh93} D. Bohm and B. J. Hiley, {\it The Undivided Universe}
               (Routledge, London and New York, 1993).
\bibitem{Hol93} P. R. Holland, {\it The Quantum Theory of Motion}
               (Cambridge University Press, Cambridge, 1993).
\bibitem{Nie00} M. A. Nielsen and I. L. Chuang, {\it Quantum Computation
                and Quantum Information} (Cambridge
                University Press, Cambridge, U.K., 2000.)
\bibitem{Lee05} Jinhyoung Lee and Seung-Woo Lee, J. Korean
                Phys. Soc. {\bf 46}, 181 (2005).
\bibitem{Ji07} Se-Wan Ji, Hai-Woong Lee, and Gui Lu Long, J. Korean
                Phys. Soc. {\bf 51}, 1245 (2007).
\bibitem {Gla63} R.\ J.\ Glauber, Phys.\ Rev.\ {\bf 130},
                2529 (1963).
\bibitem{Yue76} H. P. Yuen, Phys. Rev. A {\bf 13}, 2226 (1976).
\bibitem{Scu97} M. O. Scully and M. S. Zubairy, {\it Quantum
                Optics} (Cambridge University Press, United
                Kingdom, 1997).
\bibitem{Che07} Yong Wook Cheong and Jinhyoung Lee, J. Korean
                Phys. Soc. {\bf 51},1513 (2007).
\bibitem{Mol97} K.\ M\o lmer, \pra {\bf 55}, 3195 (1997).
\bibitem{Ben93} C. H. Bennett, G. Brassard, C. Crepeau, R. Jozsa,
               A. Peres, and W. K. Wootters,
               Phys. Rev. Lett. {\bf 70}, 1895 (1993).
\bibitem {Fur98} A.\ Furusawa, J. L. S\o rensen, S. L. Braunstein,
                 C. A. Fuchs, H. J. Kimble, and E. S. Polzik,
                 Science {\bf 282}, 706 (1998).
\bibitem{Vai94} L. Vaidman, Phys. Rev. A {\bf 49}, 1473 (1994).
\bibitem{Bra98} S. L. Braunstein and H. J. Kimble, Phys. Rev.
                Lett. {\bf 80}, 869 (1998).
\bibitem{Rud01} T. Rudolph and B. C. Sanders,
                Phys. Rev. Lett. {\bf 87}, 077903 (2001).
\bibitem{Enk02} S. J. van Enk and C. A. Fuchs,
                Phys. Rev. Lett. {\bf 88}, 027902 (2002).
\bibitem{San03} B. C. Sanders, S. D. Bartlett, T. Rudolph, and K.
                Knight, Phys. Rev. A {\bf 68}, 042329 (2003).
\bibitem{Nem04} K. Nemoto and S. Braunstein, Phys. Lett. A
                {\bf 333}, 378 (2004).
\bibitem{Wis03} H. M. Wiseman, J. Mod. Opt. {\bf 50}, 1797 (2003).
\bibitem{Smo04} J. A. Smolin, quant-ph/0407009.
\bibitem{Yue02} H. P. Yuen, a special talk in Quantum Communication, Measurement and
                Computing, July 22, 2002, Massachusetts, USA.
\bibitem{Enk05} I thank S. J. van Enk for pointing out this fact.
\bibitem{Scu99} M. O. Scully and M. S. Zubairy, {\it Quantum Optics}
                (Cambridge University Press, UK, 1999).
\bibitem{Ban98} K. Banaszek and K. Wodkiewicz, Phys. Rev. A
                {\bf 58}, 4345 (1998).
\bibitem{Jeo00} H. Jeong, J. Lee, and M. S. Kim, Phys. Rev. A
                {\bf 61}, 052101 (2000).
\bibitem{Kok00} P. Kok and S. L. Braunstein, \pra {\bf 61},
                042304 (2000).

%\vspace{5cm}

%\bibitem{diek} D. Dieks, Phys. Lett. A {\bf92}, 271 (1982).
%\bibitem{woot} W.K. Wootters and W.H. Zurek, Nature  {\bf299},
%                802  (1982).
%\bibitem{yuen} H.P. Yuen, Phys. Lett. A {\bf 113}, 405 (1986).

\end{references}
\end{document}